\documentclass[aps,pra,reprint,superscriptaddress,fleqn]{revtex4-1}

\usepackage{amsmath, amsfonts, amssymb, bm, accents, mathptmx, graphicx, siunitx,xparse, physics}
\usepackage[colorlinks=true,linkcolor=blue,anchorcolor = blue,citecolor=blue,urlcolor=blue,bookmarks=false,hypertexnames=false,pdfauthor={Abdelsalam H. M. Abdelaziz and Amarendra K. Sarma},pdfusetitle]{hyperref}
\usepackage{bookmark}
\usepackage[caption=false,listofformat=parens]{subfig}
\usepackage{romannum}

\begin{document}
\title{\textbf{Coherent Control of Hyperfine States and Slow and Fast Light Switching in Ultracold Atoms}}
\author{Abdelsalam. H. M. Abdelaziz} 
\email{ahm.abdelaziz@iitg.ac.in}
\affiliation{Department of Physics, Indian Institute of Technology Guwahati, Guwahati 781039, Assam, India}
\affiliation{Department of Physics, Faculty of Science, Suez Canal University, Ismailia 41522, Egypt}
\author{Amarendra K. Sarma} 
\email{aksarma@iitg.ac.in}
\affiliation{Department of Physics, Indian Institute of Technology Guwahati, Guwahati 781039, Assam, India}

\begin{abstract}
We propose a scheme for attaining slow and fast light via coherent control of the hyperfine ground and excited states of an ultracold atomic system. The proposed scheme is theoretically analyzed for the $D_1$ transition of ultracold $^{23}$ Na atoms. The role of field detuning and field intensity on both the population transfer and induced polarization is investigated. It is shown that  slow and fast light, with large bandwidth could be obtained in the proposed system. It is inferred that both slow and fast light could co-propagate, pertaining to different optical and Raman transitions. Further, it is observed that switching between the slow and the fast light is achievable by controlling the field intensity, as well as the field detuning. 

\end{abstract}

\maketitle

\section{\label{sec1}Introduction} 
The possibility to control the group velocity of light pulses propagating through atomic medium has led to the emergence of slow and fast light. These effects are at the forefront of optical science research due to its promising applications in optical telecommunication, interferometry, and laser radar \cite{sharping2005alloptical,sharping2005wide,boyd2006applications,residori2008slow,boyd2009controlling,boyd2009slow}. Apart from atomic media, slow and fast light phenomena is a topic of intense research in a variety of media, including optomechanics \cite{safavi-naeini2011electromagnetically,liao2020transparency}, optical fibers \cite{thevenaz2008slow} and photorefractive crystals \cite{podivilov2003light,bouldja2019improved}. In fact, recently, slow and fast light is experimentally demonstated even in a plasma medium \cite{goyon2021slow}.
Furthermore, storage and retrieval of optical pulse was reported, and recently implemented with a nonlinear metamaterial; where the concept of dark-state polariton, a combined atomic and optical excitation, clarifies how the quantum state of the probe is `imprinted' in the atoms \cite{liu2001observation,phillips2001storage,lukin2001controlling,milonni2005fast,nakanishi2018storage}. On the other hand, fast light in a medium with gain doublet, dip in gain profile, was proposed \cite{steinberg1994dispersionless}, then demonstrated by creating a Raman gain doublet using a bichromatic field \cite{wang2000gainassisted}. Moreover, fast light was demonstrated by rendering a Raman gain peak into a doublet using a coherent control field \cite{agarwal2004superluminal}. Slow and fast light have been observed extensively in solid-state materials and realized in optical fibres utilizing coherent population oscillation, stimulated Brillouin and Raman scattering \cite{bigelow2003observation,bigelow2003superluminal,thevenaz2008slow,residori2008slow,boyd2009controlling,boyd2009slow}. 

\par Slow light, also known as subluminality, stands for light propagation with a group velocity $v_g$ less than the velocity of light in vacuum $c$, while fast light, also known as superluminality, refers to light propagation with $v_g>c$ or $v_g<0$. It was thought that superluminal velocities are generally unphysical in the sense that light wave traveling with superluminal velocity does not transmit any information through the medium otherwise the causality principle of relativity theory would be violated  \cite{fitzpatrick2013oscillations}. Accordingly, subluminal group velocity can be regarded as signal velocity, which is not the case for superluminal group velocity \cite{brillouin1960wave,akulshin2010fast}.\par The group velocity for a traveling optical pulse $v_g=c/n_g$, where the group index $n_g=n+\omega\dv*{n}{\omega}$, depend on the refractive index $n$ and its dependence on frequency $\omega$\textemdash the dispersion $\dv*{n}{\omega}$. It is well known, from Kramers–Kronig relations which satisfy causality principle, that the dispersion changes sign steeply about absorption line (or closely-spaced absorption doublet), equivalent to dip in gain profile; where the steep anomalous (negative) dispersion near absorption line center results in fast light and the normal (positive) dispersion in the wings of absorption line results in slow light; likewise, slow light, associated with normal  dispersion, is expected near the center of gain line (or closely-spaced gain doublet), equivalent to dip in absorption profile which is also known as spectral hole, and fast light, associated with anomalous dispersion, is expected in the wings of gain line \cite{brillouin1960wave,garrett1970propagation,chiao1993superluminal,bolda1994optical,stenner2005distortion,milonni2005fast,boyd2007slow,boyd2009slow}. The observation of electromagnetically induced transparency (EIT) \cite{boller1991observation,harris1992dispersive}, a quantum destructive interference effects in which a transparency window is generated between two absorption peaks, a dip in absorption profile, opens the way for slow light demonstration in Bose-Einstein condensate of  sodium atoms with $v_g= 17~\si{\meter\per\second}$  \cite{hau1999light}, and the observation of $v_g= 90~\si{\meter\per\second}$ \cite{kash1999ultraslow} and $8~\si{\meter\per\second}$ \cite{budker1999nonlinear} in rubidium vapor. Despite the near complete symmetry between slow and fast light effects, it was found that large fractional, delays is easily attainable than advances, on resonance. This may be attributed to the constraints imposed by both the maximum allowable loss or gain in a medium. The loss should not be too large, as the transmitted pulse would be too weak to be noticeable. Again, the gain should not also be too large at any frequency; otherwise the process of amplified spontaneous emission may occur resulting in the depletion of gain of the material\cite{cao2003negative,boyd2007slow}. Slow light of EIT medium can be switched to fast light by driving the transition between the ground states using an additional control field in a closed system \cite{agarwal2001knob}, or by tuning the phase of one of the fields in an open system \cite{sahrai2004tunable}; conversely, fast light of single absorption line can be switched to slow light by controlling the coupling field intensity \cite{kim2003observation}. Copropagating slow and fast light manifested by EIT-assisted nonlinear gain and absorption has been observed experimentally \cite{zhang2006copropagating}. In this paper, we investigate the role of field detuning on both the population transfer and the induced polarization for optical and Raman transitions of the system. we observe slow and fast light, with large bandwidth at intermediate field intensity. The paper is organized as follows: In Sec. \ref{sec2} we present the density matrix equations for the proposed four-level atomic system driven by a single coherent laser field. Section \ref{sec3} contains our simulated results and discuss population transfer, coherence, and slow and fast light, with large bandwidth, followed by conclusions in Sec. \ref{sec4}.

\section{\label{sec2}The Model}
\begin{figure}
\subfloat{
	\begin{minipage}[t]{1\linewidth}
		\centering
		\includegraphics[width=1\linewidth]{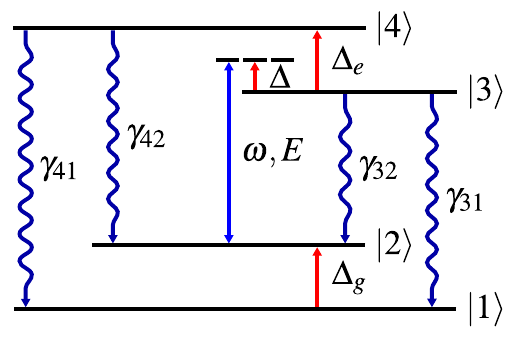}\label{1}
	\end{minipage}}
\caption{(Color online)\label{fig1} Schematic of the Four-level atomic system.}
\end{figure}

We consider a four-level atomic system, where two hyperfine ground states, $\ket{1}$ and $\ket{2}$, are coupled by a single coherent laser field of frequency $\omega$ and amplitude $E$ to two hyperfine excited states, $\ket{3}$ and $\ket{4}$, as depicted in Fig. \ref{fig1}. This system can be realized by the hyperfine states associated with the $D_1$ transition of an alkali atom with a nuclear spin quantum number $I=3/2$ ($^7$Li, $^{23}$Na, $^{39,\,41}$K, and $^{87}$Rb), where the total angular momenta of the hyperfine ground and excited states are $F=1,\,2$ and $F^\prime=3,\,4$, respectively. The electric field can be written as:
$\mathbf{E}=(\vu{\boldsymbol{\epsilon}}Ee^{i\omega t}+c.c)/2$.
Under the dipole and the rotating-wave approximations, the Hamiltonian of a four-level atomic system is given by:
\begin{equation}
\mathbf{H}/\hbar=
\begin{bmatrix}
0 &0 &\Omega_{13}&\Omega_{14}\\ 
0&\Delta_g&\Omega_{23} &\Omega_{24}\\ 
\Omega_{13}&\Omega_{23}&-\Delta&0\\
\Omega_{14}&\Omega_{24}&0&\Delta_e-\Delta
\end{bmatrix}\;,
\end{equation}
where $\Omega_{ij}=-\mu_{ij}E/2\hbar$ is one-half Rabi frequency for the transition with electric dipole moment $\mu_{ij}$. 
$\Delta_g=\omega_{21}$ and $\Delta_e=\omega_{43}$ are ground and excited states hyperfine energy splittings, respectively. $\Delta=\omega-\omega_{32}$ is the laser field detuning. It should be noted that $\omega_{ij}=\omega_i-\omega_j$. The density matrix equations describing the atomic system dynamics are given by:
\begin{equation}\label{eq2}
	\begin{aligned} 
		\dot{\rho}_{11}&=  \gamma_{31}\rho_{33} + \gamma_{41}\rho_{44} + i[\Omega_{13}\rho_{13}  + \Omega_{14}\rho_{14}-c.c]\\
		\dot{\rho}_{22}&= \gamma_{32}\rho_{33} + \gamma_{42}\rho_{44} + i[\Omega_{23}\rho_{23} + \Omega_{24}\rho_{24}-c.c]\\
		\dot{\rho}_{33}&= -(\gamma_{31} + \gamma_{32})\rho_{33} - i[\Omega_{13}\rho_{13}  + \Omega_{23}\rho_{23}-c.c]\\
		\dot{\rho}_{44}&= -(\gamma_{41} + \gamma_{42})\rho_{44} - i[\Omega_{14}\rho_{14}  + \Omega_{24}\rho_{24}-c.c]\\
		\dot{\rho}_{31}&=  \Gamma_{31}\rho_{31} - i[\Omega_{13}(\rho_{11} - \rho_{33}) - \Omega_{14}\rho_{34} + \Omega_{23}\rho_{21}]\\
		\dot{\rho}_{32}&= \Gamma_{32}\rho_{32} - i[\Omega_{13}\rho_{12} + \Omega_{23}(\rho_{22} - \rho_{33}) - \Omega_{24}\rho_{34}]\\
		\dot{\rho}_{41}&= \Gamma_{41}\rho_{41} + i[\Omega_{13}\rho_{43} - \Omega_{14}(\rho_{11} - \rho_{44}) - \Omega_{24}\rho_{21}]\\
		\dot{\rho}_{42}&= \Gamma_{42}\rho_{42} - i[\Omega_{14}\rho_{12} - \Omega_{23}\rho_{43} + \Omega_{24}(\rho_{22} - \rho_{44})]\\
		\dot{\rho}_{21}&=  \Gamma_{21}\rho_{21}+i[\Omega_{13}\rho_{23} + \Omega_{14}\rho_{24} - \Omega_{23}\rho_{31} - \Omega_{24}\rho_{41}]\\
		\dot{\rho}_{43}&= \Gamma_{43}\rho_{43} + i[\Omega_{13}\rho_{41} - \Omega_{14}\rho_{13} + \Omega_{23}\rho_{42} - \Omega_{24}\rho_{23}]
	\end{aligned}\;,
\end{equation} 
with $\rho_{ij}=\rho_{ji}^\ast$ and $\sum_{i=1}^{4}\rho_{ii}=1$. $\gamma_{ij} $ denotes the spontaneous decay rate from state $\ket{i}$ to $\ket{j}$. 
$\Gamma_{21}=-i\Delta_g$, $\Gamma_{31}=i(\Delta - \Delta_g)-(\gamma_{31} + \gamma_{32})/2$, $\Gamma_{32}=i\Delta-(\gamma_{31} + \gamma_{32})/2$, $\Gamma_{41}=i(\Delta-\Delta_g - \Delta_e )-(\gamma_{41} + \gamma_{42})/2$,  $\Gamma_{42}=i(\Delta - \Delta_e)-(\gamma_{41} + \gamma_{42})/2$ and $\Gamma_{43}=-i\Delta_e -(\gamma_{31} + \gamma_{32} + \gamma_{41} + \gamma_{42})/2$. As we are dealing with dense media, in which many atoms exist within a cubic resonance wavelength, the near-dipole-dipole interaction (NDD) needs to be considered by incorporating local-field correction resulting in the following relation:  $\Omega_{ij}=\Omega-\epsilon_{ij} \,Re(\rho_{ij})$, where $\epsilon_{ij}=\frac{N\mu_{ij}^2}{3\epsilon_0\hbar}$ provided that $N$ is the atomic number density and $\epsilon_0$ is the permittivity of free space \cite{bowden1993dipoledipole,xia2005dipoledipole}. We choose the average of ground and excited states hyperfine energy splittings $\Delta_u=(\Delta_g+\Delta_e)/2$ as a unit scale for detuning, and define a new detuning $\Delta_c=\Delta-\Delta_u$ by raising the point of zero detuning by $\Delta_u$.

It may be useful to be reminded that the atomic system response to the applied fields is determined by the complex susceptibility parameter, $\chi=\chi_R+i\chi_I$, which is connected to $\rho_{ij}$, which is again a complex quantity, via the following relation \cite{boyd2020nonlinear,abdelaziz2020effective}:
\begin{equation}
	\chi_{ij}=\frac{N\mu_{ij}}{\epsilon_oE}\rho_{ij}=-\frac{N\abs{\mu_{ij}}}{\epsilon_oE}\rho_{ij}\;.
\end{equation}

It is well known that the real part of susceptibility, i.e. $\chi_R$ (which is again related to $Re(\rho_{ij})$), is related to dispersion while the imaginary one, i.e. $\chi_I$ (which in turn is associated to $Im(\rho_{ij})$), is related to absorption of the atomic system \cite{boyd2020nonlinear}. 

\section{\label{sec3}Results and Discussions}
\begin{figure*}
\subfloat{
	\begin{minipage}[t]{0.5\linewidth}
		\centering
		\includegraphics[width=1\linewidth]{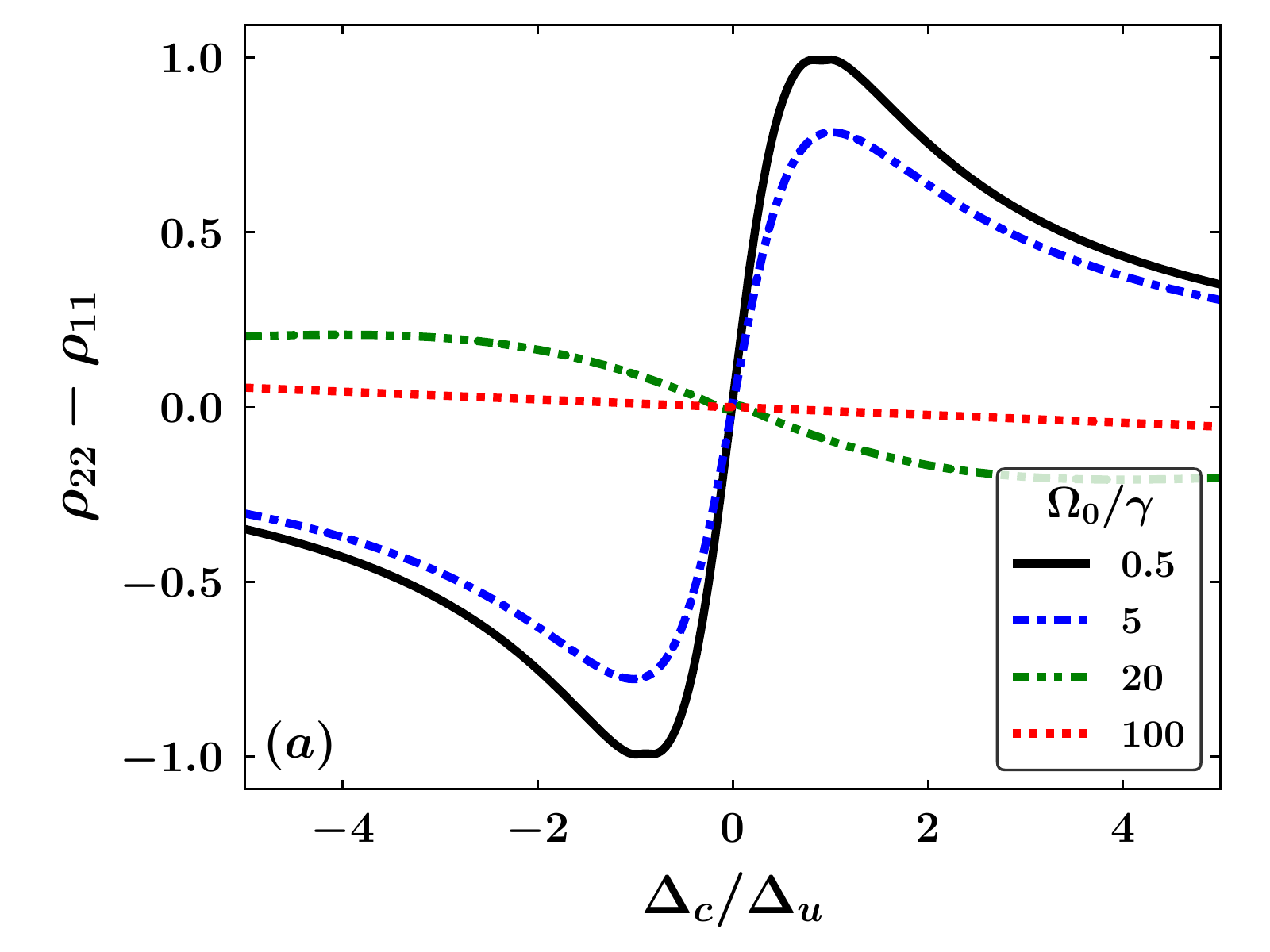}\label{2a}
	\end{minipage}}
\subfloat{
	\begin{minipage}[t]{0.5\linewidth}
		\centering
		\includegraphics[width=1\linewidth]{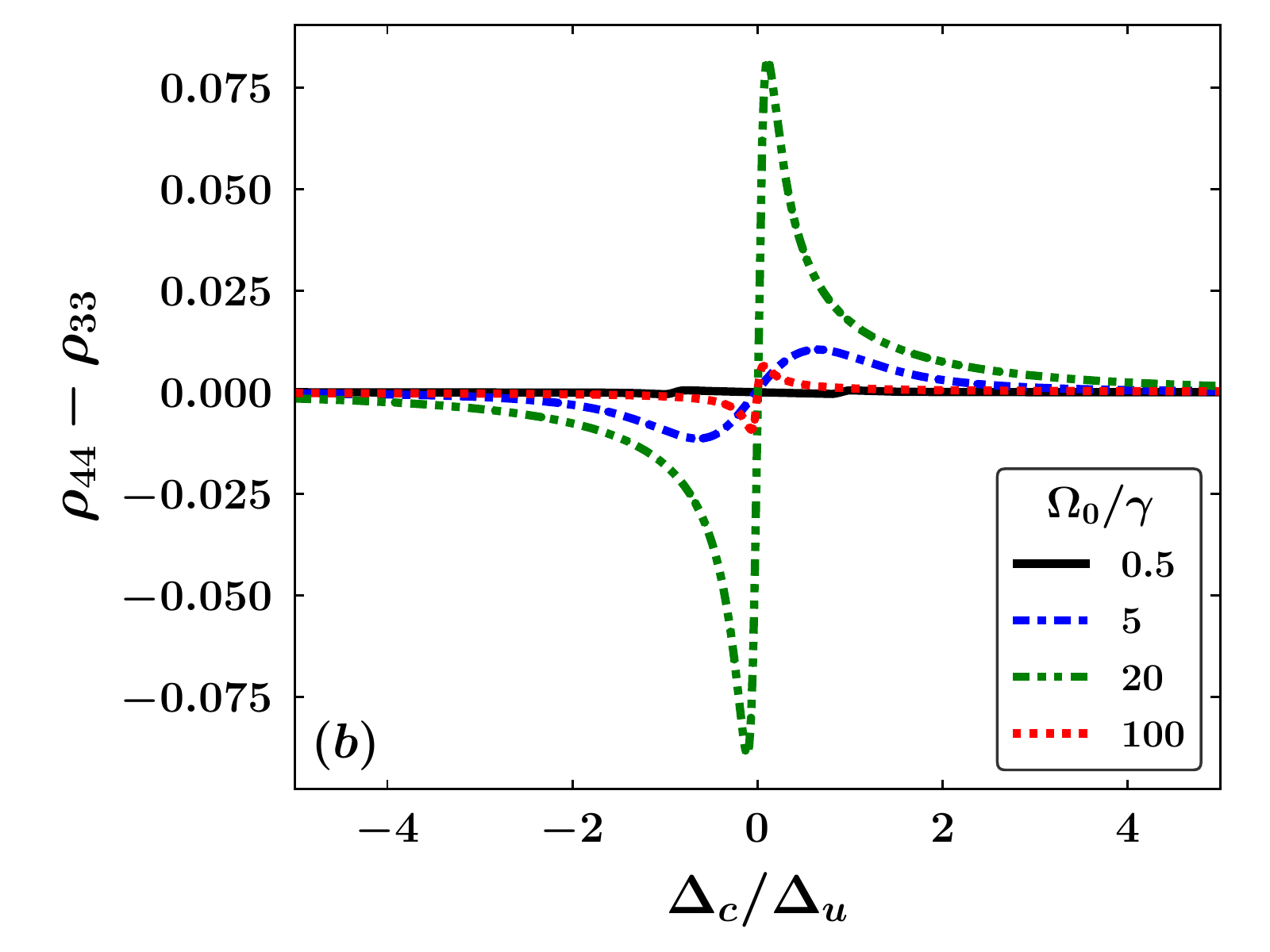}\label{2b}
	\end{minipage}}
\caption{(Color online)\label{fig2} Population transfer as a function of field detuning for the Raman transition between the two hyperfine: (a) ground-, (b) excited- states.}
\end{figure*}

We solve Eqs. \ref{eq2} and their complex conjugates at the steady state to investigate the role of field detuning on both the population transfer and the induced polarization for optical and Raman transitions of the system. We consider the $D_1$ transition of Bose-Einstein condensate of $^{23}$Na atoms with atomic number density N of $1.5\times 10^{20}~\si{\meter^{-3}}$ at a temperature of $2.0~\si{\micro\kelvin}$ \cite{davis1995boseeinstein}. Note that all hyperfine transitions have the same, spontaneous decay rate $\gamma/{2\pi }$ of $9.76~\si{\mega\hertz}$ and electric dipole moment $\mu$ of $21.1165 \times 10^{-30}~\si{\coulomb\cdot\meter}$ \cite{sansonetti2008wavelengthsa,steck2019sodium,sydoryk2008broadening}.  The ground and excited states hyperfine energy splittings are $\Delta_g/{2\pi}=1771.62$ and $\Delta_e/{2\pi }=188.88~\si{\mega\hertz}$, respectively. Accordingly, the unit scale for detuning is $\Delta_u/{2\pi}=980.25~\si{\mega\hertz}$. Figure \ref{fig2} exhibits the role of field detuning on population transfer for four different field intensities, namely $0.5\gamma$, $5\gamma$, $20\gamma$, and $100\gamma$ which would be referred respectively as small, low-intermediate, high-intermediate, and large field intensities throughout the rest of the paper. Interestingly, we observe complete population transfer, for Raman transition, between the two hyperfine ground states [Fig. \subref{2a}] at small field intensity around the positive and negative peaks of $\Delta_c/\Delta_u\approx\pm1$, i.e., the field is near resonant between $\ket{4}$ and $\ket{1}$ states, and $\ket{3}$ and $\ket{2}$ states respectively. As the field intensity is increased, the curve depicting population transfer gets flatten gradually leading to insignificant population transfer for large field intensity. Then the curve reverses its direction near about $\Delta_c=0$ at high-intermediate field intensity of $20\gamma$. For the other Raman transition between the two hyperfine excited states [Fig. \subref{2b}], partial population transfer occurs at intermediate field intensities with positive and negative peaks about $\Delta_c=0$. It can be seen that there is insignificant population transfer for small and large field intensities. In view of these results, it could be concluded that population inversion for optical transitions does not occur.

\begin{figure*}
\subfloat{
	\begin{minipage}[t]{0.5\linewidth}
		\centering
		\includegraphics[width=1\linewidth]{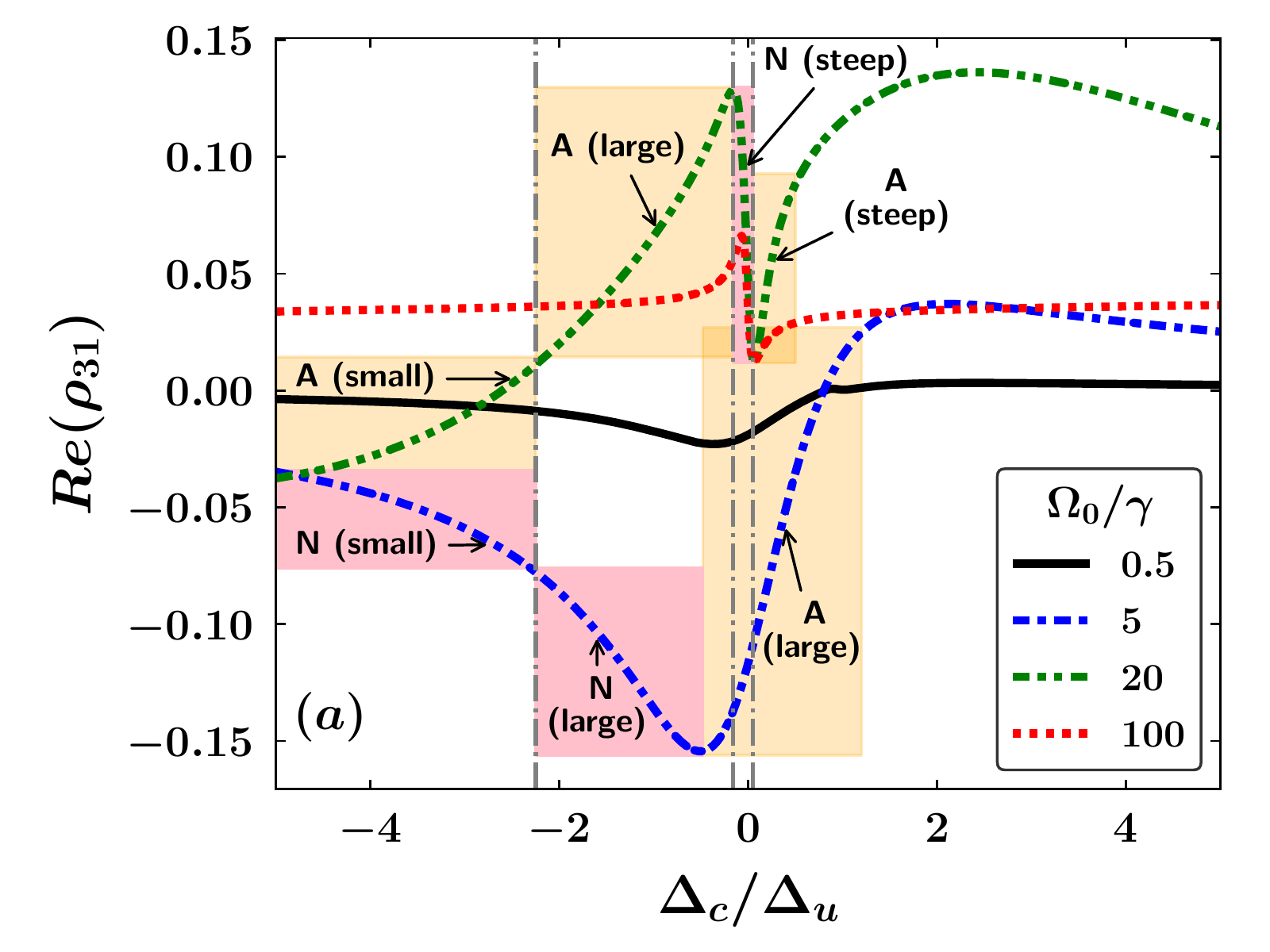}\label{3a}
	\end{minipage}}
\subfloat{
	\begin{minipage}[t]{0.5\linewidth}
		\centering
		\includegraphics[width=1\linewidth]{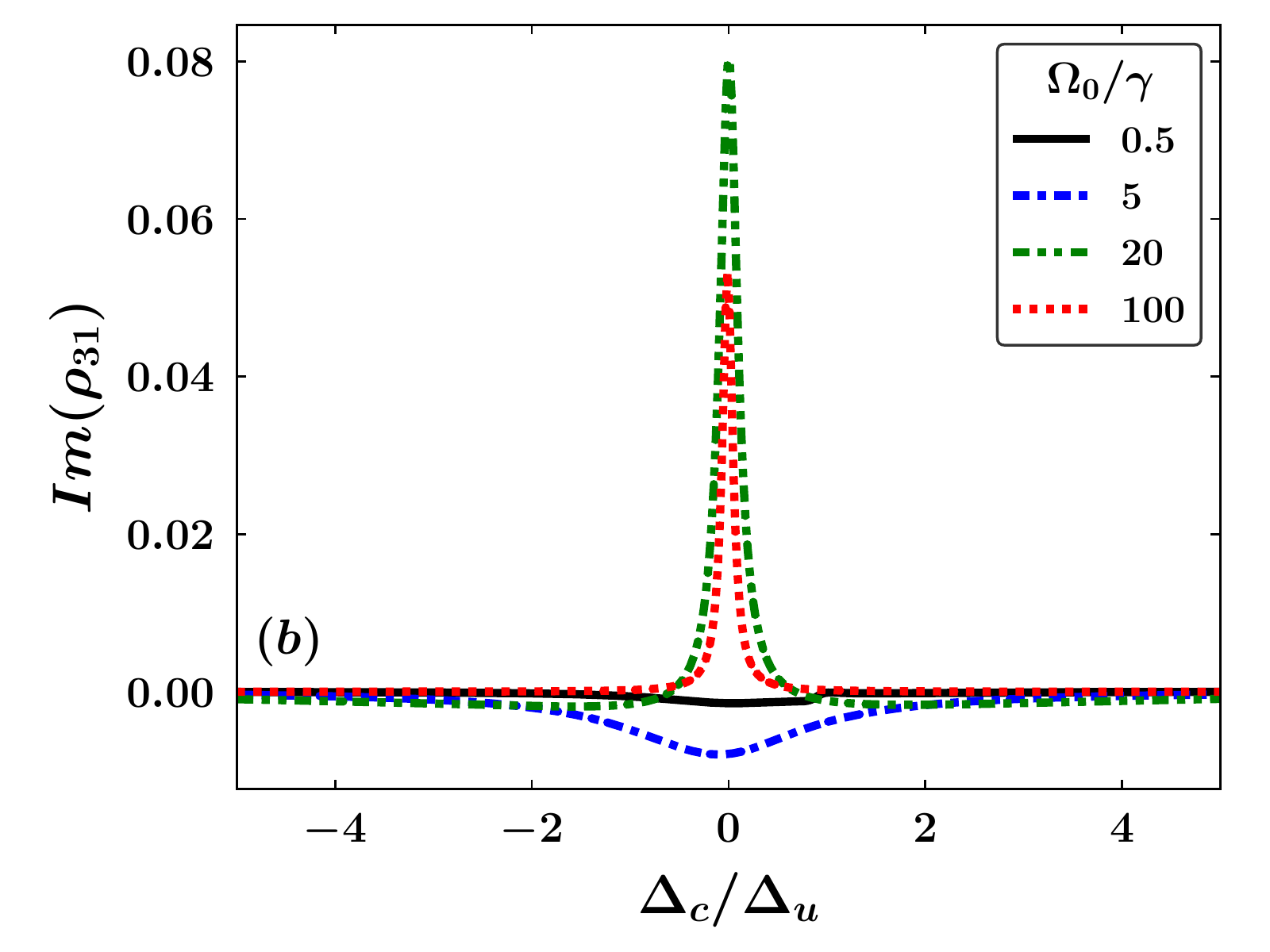}\label{3b}
\end{minipage}}

\subfloat{
	\begin{minipage}[t]{0.5\linewidth}
		\centering
		\includegraphics[width=1\linewidth]{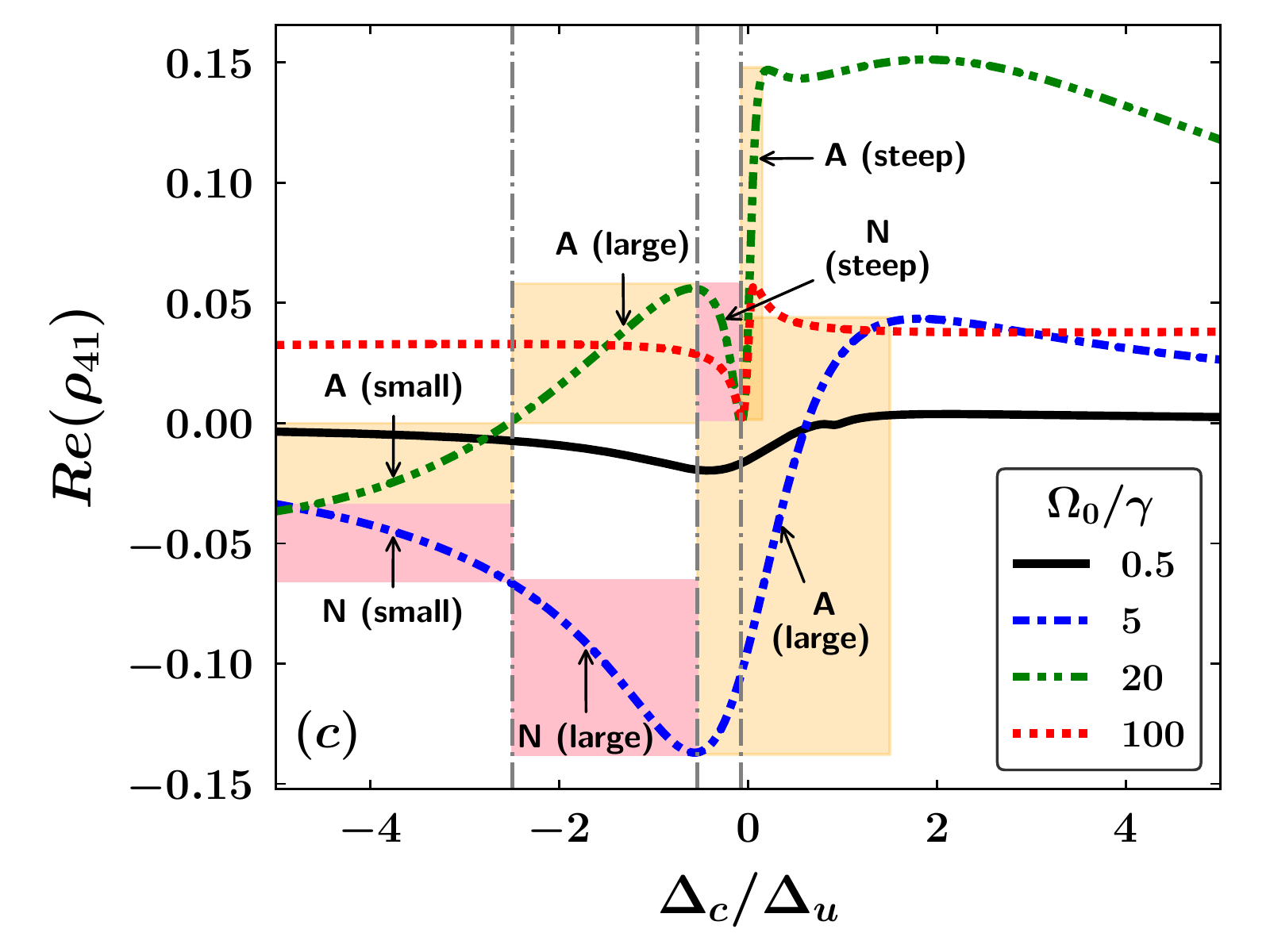}\label{3c}
	\end{minipage}}
\subfloat{
	\begin{minipage}[t]{0.5\linewidth}
		\centering
		\includegraphics[width=1\linewidth]{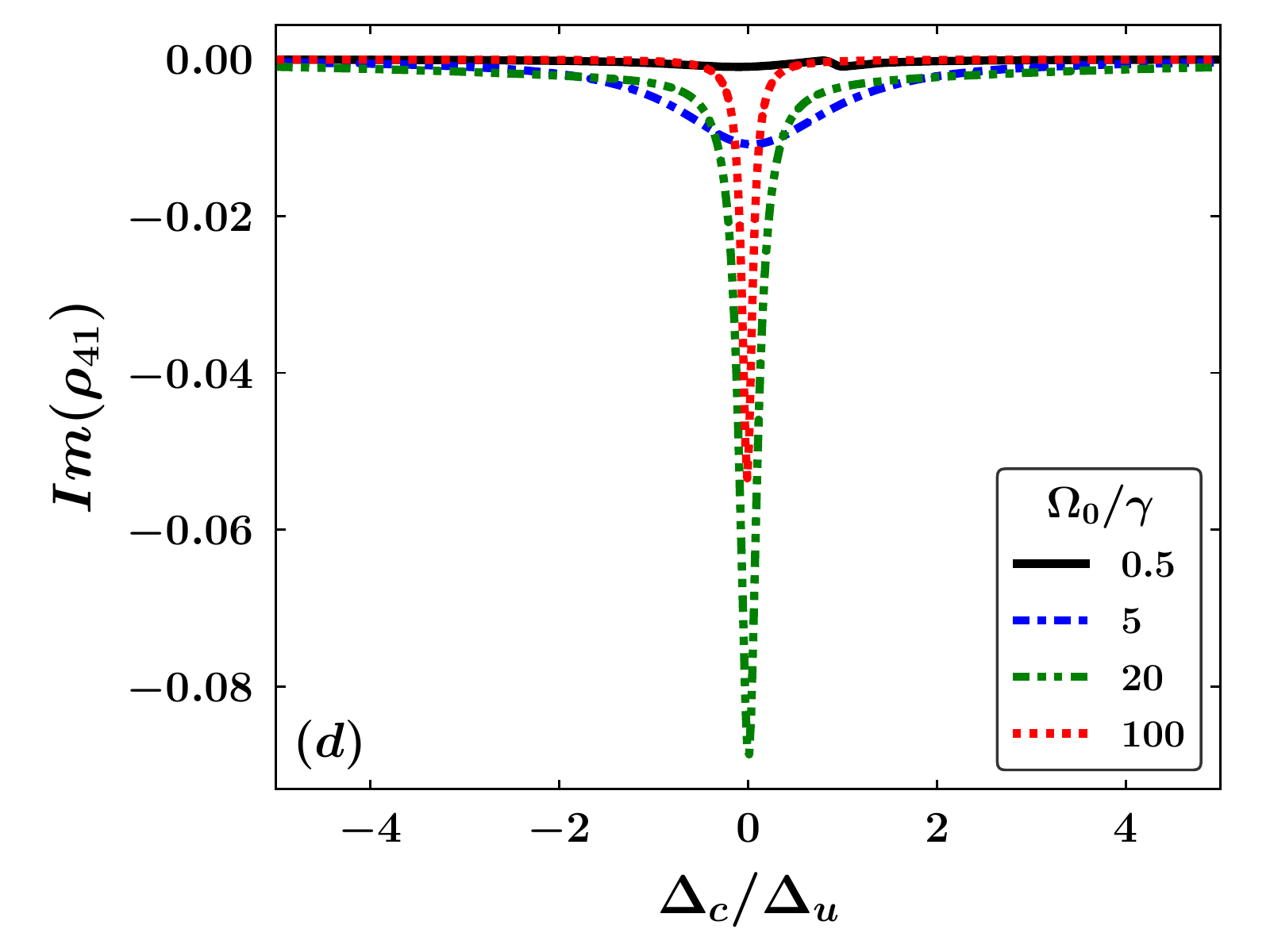}\label{3d}
\end{minipage}}
\caption{(Color online)\label{fig3} Spectra of the real and imaginary components of polarization for the optical transitions between 
$\ket{3}\leftrightarrow\ket{1}$ [(a) and (b)] and $\ket{4}\leftrightarrow\ket{1}$ [(c) and (d)]. $\bf N$ and $\bf A$ refers to 'Normal' and 'Anamolous' dispersion respectively.}
\end{figure*}

In Fig. \ref{fig3}, we depict the effect of field detuning on the real and imaginary components of the induced polarization, which are correlated to dispersion and absorption respectively, for optical transitions, $\ket{3}\leftrightarrow\ket{1}$ and 
$\ket{4}\leftrightarrow\ket{1}$, under different field intensities. In the rest of the manuscript, for brevity and clarity of explanations, we refer the optical transitions, $\ket{3}\leftrightarrow\ket{1}$ and $\ket{4}\leftrightarrow\ket{1}$ as optical transitions \Romannum{1} and \Romannum{2} respectively. As could be seen from Fig.\ref{fig3}, the spectral range of the region of anomalous dispersion associated with the absorption line for the optical transition \Romannum{1} and \Romannum{2} at low-intermediate field intensity is approximately $2\ \Delta_u$. This region of (linear) anomalous dispersion is appropriate for realizing fast light with large bandwidth. On the other hand, the region of normal dispersion of the red-detuned wing of absorption line, corresponding to intermediate field intensity, has a spectral range of approximately $4.5\ \Delta_u$. This region could be split into two sub-regions of (nearly linear) small and large normal dispersion. The latter sub-region is appropriate for realizing slow light with large bandwidth. Interestingly, the large-linewidth absorption-profile changes to gain profile, for the optical transition \Romannum{1} [Figs. \subref{3a} and \subref{3b}] at the high-intermediate field intensity. Now, the spectral range of the region of steep-normal dispersion associated with the gain line is found to be approximately $0.15\ \Delta_u$. This region of dispersion, which is nearly linear, is appropriate for realizing slow light. It could be seen that the red-detuned wing of the gain line, which has a spectral range of approximately $4.9\ \Delta_u$, exhibits anomalous dispersion. This region could be split into two sub-regions of (nearly linear) small and large normal dispersion. The latter sub-region is appropriate for realizing fast light with large bandwidth. Again, it could be observed that there is a steep but small anomalous dispersion regime, near $\Delta_c=0$ and $\Delta_c>0$, that could be utilized for attaining fast-light. Thus, by controlling the field intensity or the field detuning, switching from slow to fast light or back, is feasible.

We can carry out a similar analysis for the optical transition \Romannum{2}. Here, we observe that the absorption profile with large-linewidth, corresponding to intermediate field intensity, [Figs. \subref{3c} and \subref{3d}] transforms into an absorption profile with narrow-linewidth at high-intermediate field intensity. The spectral range of the region of  very steep-anomalous dispersion associated with the absorption line is approximately $0.15\ \Delta_u$. This  linear dispersion region, near $\Delta_c=0$ and $\Delta_c>0$, is appropriate for realizing fast light. One could observe a steep normal dispersion region at the other side, i.e. near $\Delta_c=0$ and $\Delta_c <0$; this region could be used for slow-light. The region of dispersion corrsponding to the red-detuned wing of the absorption line has a spectral range of approximately $4\ \Delta_u$. This region could be further divided into two sub-regions of (nearly linear) small-anomalous and large-anomalous as indicated in Fig. \subref{3c}. The latter sub-region is appropriate for realizing fast light with large bandwidth. Thus, by tuning the field intensity or the frequency of the external field, it is possible to attain switching from slow to the fast light or back. One of the advantages of the proposed scheme is that the switching mechanisms could be realized with large bandwidth for both the optical transitions. 
	
In addition to these switching behaviors from the slow to the fast light or the reverse, it could be seen from Fig. \ref{fig3} that, at high-intermediate field intensity, the slow light associated with optical transition \Romannum{1} could co-propagate with the fast light related with the optical transition \Romannum{2}. In passing, it is worthwhile to mention that the spectra of polarization for the optical transitions,
$\ket{4}\leftrightarrow\ket{2}$ and $\ket{3}\leftrightarrow\ket{2}$  are not depicted as we find that these transitions are connected to optical transitions \Romannum{1} and \Romannum{2}, by dint of the following empirical relations:   
\begin{equation}\label{eq3}
	\begin{aligned} 
		\rho_{42}(\pm\Delta_c)+\rho^\ast_{31}(\mp\Delta_c)&=0\\
		\rho_{32}(\pm\Delta_c)+\rho^\ast_{41}(\mp\Delta_c)&=0
	\end{aligned}\
\end{equation}

We can also make few observations on maximum attainable coherences in the proposed system. It could be seen from Fig. \subref{3a} that for the transition \Romannum{1}, we can obtain a coherence of $\approx$ 0.3 (on the scale of maximum attainable coherence of 0.5) at the low-intermediate intensity. On the other hand, the same is feasible at low- as well as, high-intermediate intensity for the transition \Romannum{2}, as could seen from Fig. \subref{3c}.  

\par Now, as regards the Raman transitions, $\ket{2}\leftrightarrow\ket{1}$ and $\ket{4}\leftrightarrow\ket{3}$, are concerned, the behavior of the corresponding induced polarizations could be predicted from the optical transitions themselves, as follows. Let us
consider the steady state solutions of the induced polarization for Raman transitions by setting $\dot{\rho}_{21}$ and $\dot{\rho}_{43}$ of Eqs. \ref{eq2} to zero; this leads to the following equations:
\begin{equation}\label{eq4}
	\begin{aligned} 	 \Gamma_{21}\rho_{21}&=-i\Omega[\rho_{23} + \rho_{24} - \rho_{31} - \rho_{41}]\\
		\Gamma_{43}\rho_{43} &=- i\Omega[\rho_{41} - \rho_{13} + \rho_{42} - \rho_{23}]
	\end{aligned}\;.
\end{equation}  

 Here, for simplicity we neglect the near dipole-dipole interaction, which may be of importance only at small field intensity.  Using Eqs. \ref{eq3}, we can express Eqs.\ref{eq4} as follows: 
\begin{equation}\label{eq5}
	\begin{aligned} 	 \Gamma_{21}\rho_{21}&=i\Omega[ \rho_{31}^R+\rho_{41}^R]\\
		\Gamma_{43}\rho_{43} &= i\Omega[\rho_{31}^{\ast R}-\rho_{41}^R]
	\end{aligned}\;,
\end{equation}  
where $\rho_{ij}^R=\rho_{ij}(\pm\Delta_c) + \rho_{ij}(\mp\Delta_c)$. 

\begin{figure*}
	\subfloat{
		\begin{minipage}[t]{0.5\linewidth}
			\centering
			\includegraphics[width=1\linewidth]{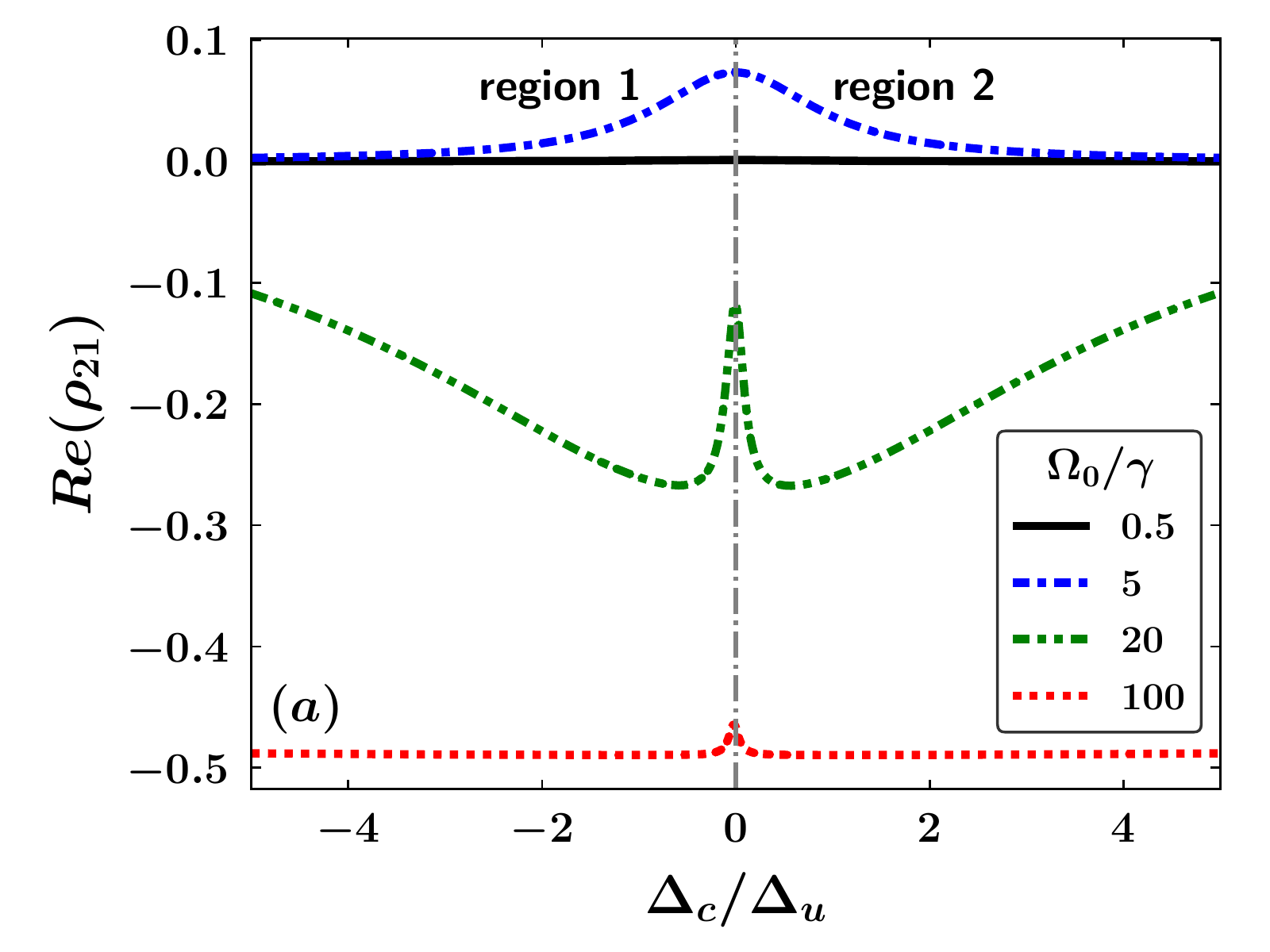}\label{4a}
	\end{minipage}}
	\subfloat{
		\begin{minipage}[t]{0.5\linewidth}
			\centering
			\includegraphics[width=1\linewidth]{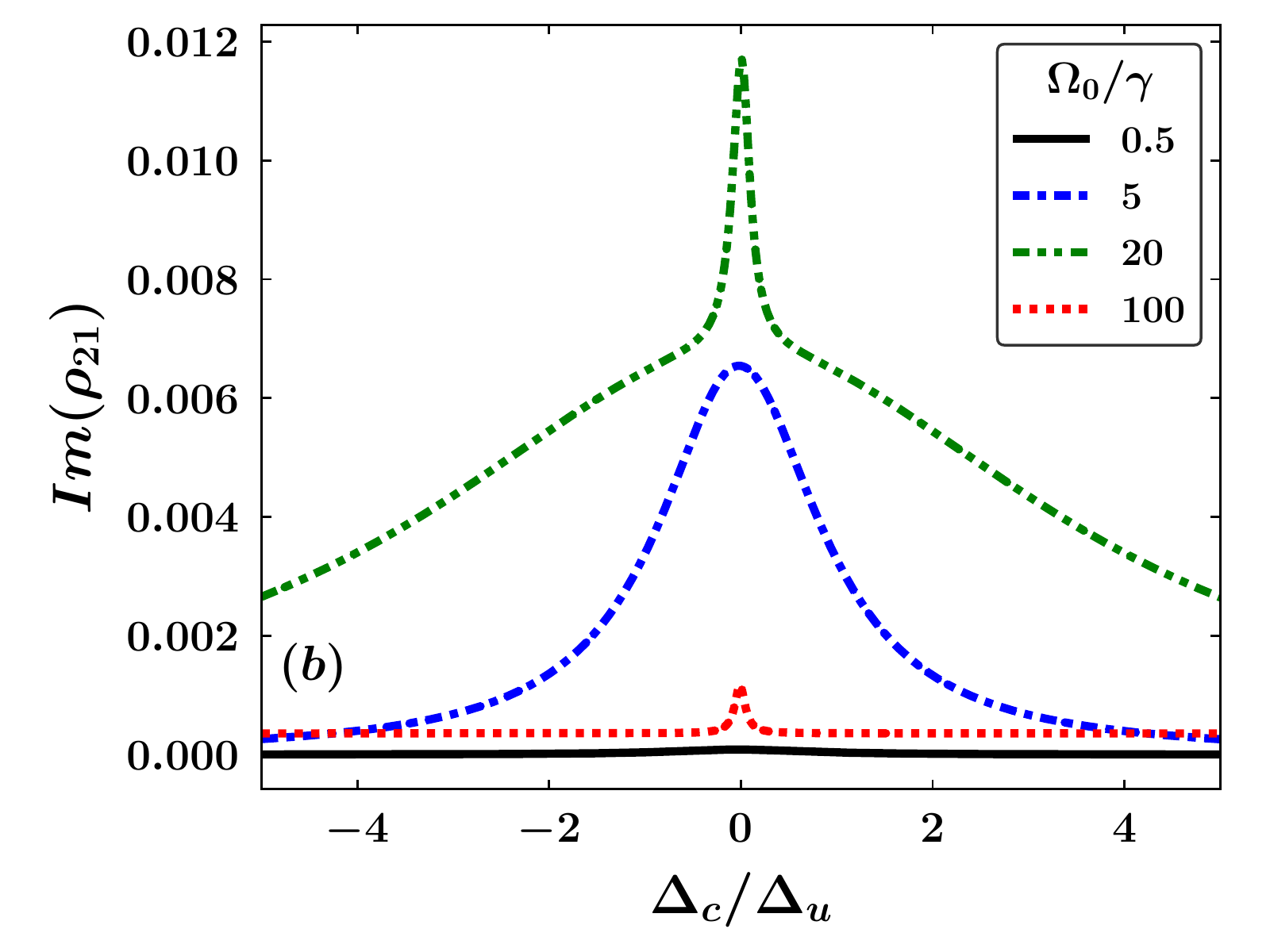}\label{4b}
	\end{minipage}}
	
	\subfloat{
		\begin{minipage}[t]{0.5\linewidth}
			\centering
			\includegraphics[width=1\linewidth]{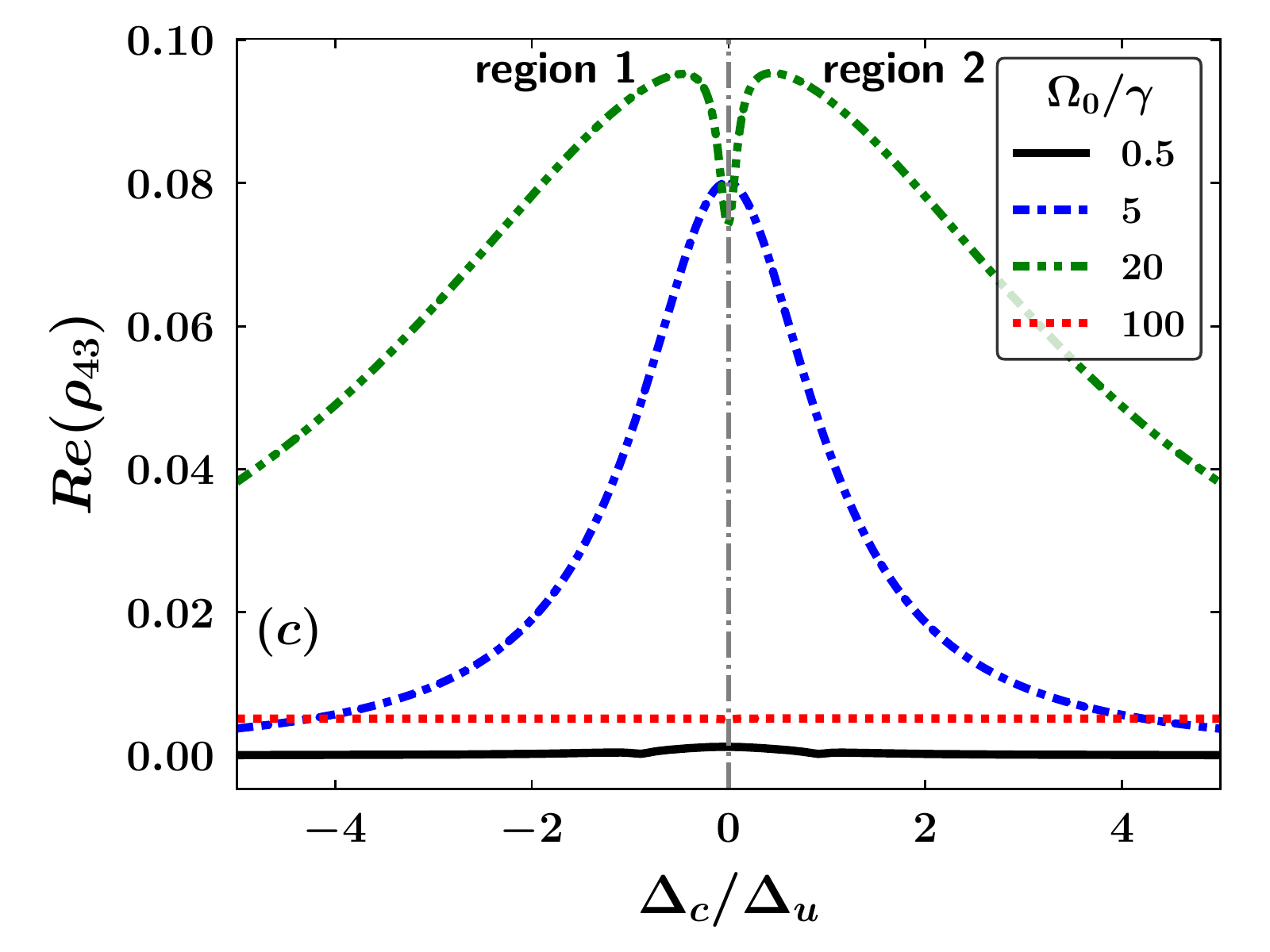}\label{4c}
	\end{minipage}}
	\subfloat{
		\begin{minipage}[t]{0.5\linewidth}
			\centering
			\includegraphics[width=1\linewidth]{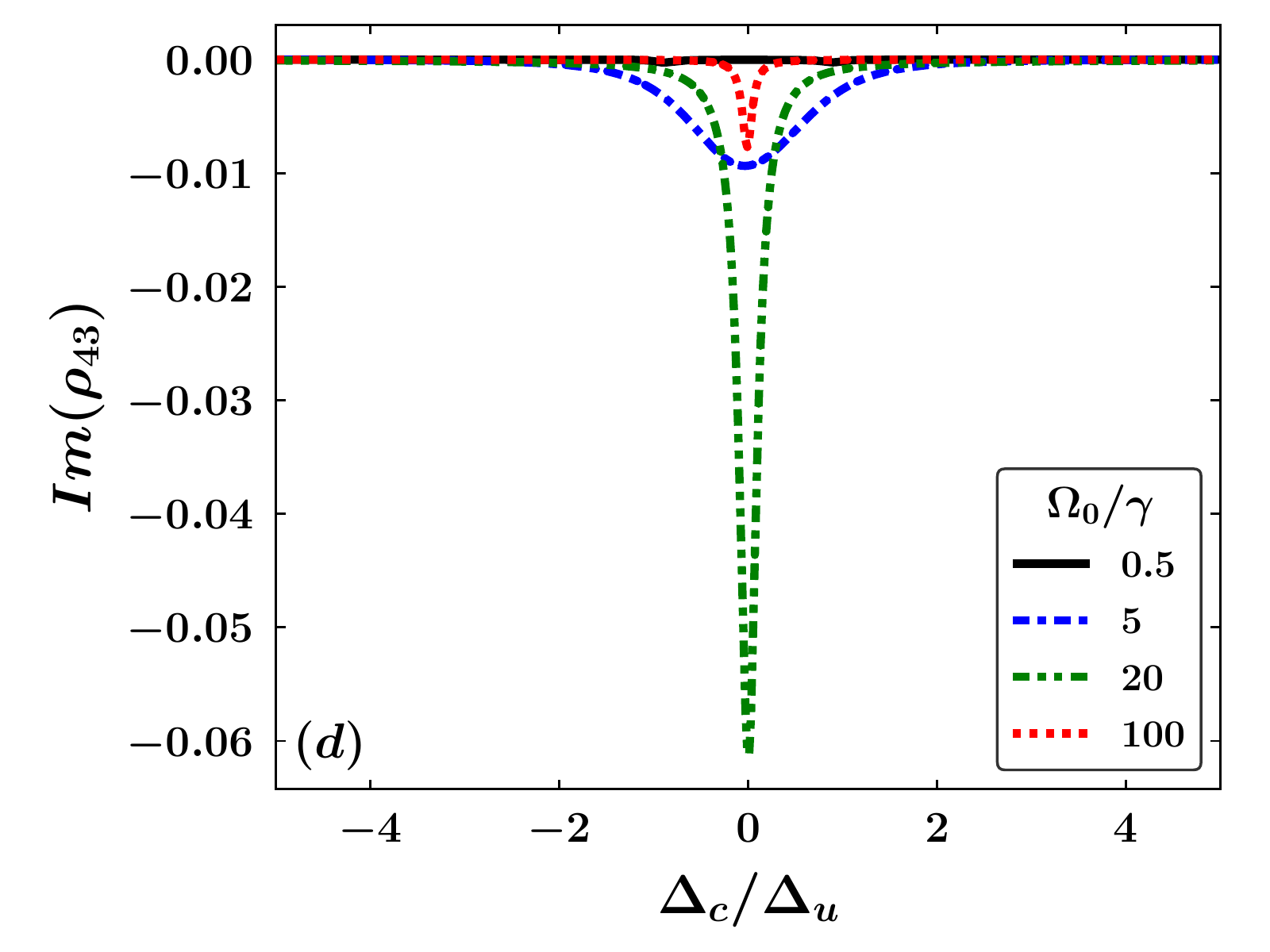}\label{4d}
	\end{minipage}}
	\caption{(Color online)\label{fig4} Spectra of the real and imaginary components of the induced polarization for the Raman transitions between the two hyperfine: (a) and (b)  ground-, (c) and (d) excited- states.}
\end{figure*}

It is easy to infer directly from Eqs. \ref{eq5}, that the spectra of the induced polarization for Raman transitions is symmetric about $\Delta_c=0$ axis, i.e. it is even function of $\Delta_c$. Bearing that in mind, we depict the effect of field detuning on the real and imaginary components of the induced polarization for the Raman transitions under different field intensity in Fig. \ref{fig4}. For brevity of explanations, we split the spectra of the induced polarization into two regions, {1} and {2} as shown in Fig.\ref{fig4}. Each region has a spectral range of $5\ \Delta_u$. 
Figs. \subref{4a} and \subref{4b} discuss the Raman transition $\ket{2}\leftrightarrow\ket{1}$, while Figs. \subref{4b} and \subref{4c} consider the one with $\ket{4}\leftrightarrow\ket{3}$.
It could be seen from Fig. \subref{4a} that, at low-intermediate intensity near $\Delta_c=0$, we have a linear anomalous dispersion region in {1} and a normal one in {2}. These regions correspond to a broad-gain profile, as could be observed from Fig. \subref{4b}. Thus just by a slight change in field detuning, one can switch from a fast to a slow light or back. It is interesting to observe that, at high-intermediate field intensity, the gain profile gets wider, and a very sharp spike appears at the gain line center. Now the corresponding dispersion curve exhibits, near $\Delta_c=0$, very steep anomalous and normal dispersion in region {1} and {2} respectively. This could again be exploited for fast and slow light applications.Thus, by controlling the field intensity, switching from fast to the slow light could be achieved. However, it should be noted that the proposed scheme is not effective at very large intensity or at small intensity, as evident from Fig.\ref{fig4}. 

 
 Now, as regards the Raman transition, $\ket{4}\leftrightarrow\ket{3}$ is concerned, one can observe steep, normal and anomalous dispersion in region 1 and 2 respectively at high-intermediate intensity near $\Delta_c=0$. In contrast, we can notice (near linear) anomalous and normal dispersion at low-intermediate intenstity in the respective regions. It is worthwhile to note that, here, these dispersion curves pertain to aborption lines; while in the case of Raman transition,$\ket{2}\leftrightarrow\ket{1}$, discussed earlier, the dispersion curves are related to gain lines. Once again, we find that either by field detuning or changing the intensity we can switch from fast to slow light or back. One could have a larger band-width, and lower absorption, in the case of low-intermediate intensity compared to the high-intermediate one. Again, for reasons raised earlier, the Raman transition involving the hyperfine ground states is preferable for exploration of fast light phenomena over that of the excited ones. Further, it could be observed that, at high-intermediate field intensity, the fast (slow) light-associated with the gain line, corresponding to $\ket{2}\leftrightarrow\ket{1}$ Raman transition could co-propagate with the slow (fast) light associated with loss-line for the Raman transition $\ket{4}\leftrightarrow\ket{3}$. Finally, we observe that, while very little coherence is possible to attain between the hyperfine excited states, it is possible to obtain maximum coherence of $\approx$ 0.5 between the two hyperfine ground states, though, at large field intensity. 
 
\section{\label{sec4}Conclusions}
In conclusion, we have theoretically analyzed the role of field detuning on both the population transfer and the induced polarization for optical and Raman transitions in an ultracold atomic system.  We have observed complete population transfer, for Raman transition, between the two hyperfine ground states at small field intensity when the field is near resonant between $\ket{4}$ and $\ket{1}$ states or $\ket{3}$ and $\ket{2}$ states. We have ascertained that slow and fast light, with large bandwidth are achievable at intermediate field intensity for the optical transitions, $\ket{3}\leftrightarrow\ket{1}$ and $\ket{4}\leftrightarrow\ket{1}$. In addition, by controlling the field intensity, switching from the slow to the fast light, with large bandwidth and the reverse are attainable. It is observed that at high-intermediate field intensity, the slow-light associated with gain for the optical transition $\ket{3}\leftrightarrow\ket{1}$ co-propagates with fast-light associated with loss for the optical transition $\ket{4}\leftrightarrow\ket{1}$. Similarly, we have observed that slow and fast light, with large bandwidth are feasible at intermediate field intensities for Raman transitions of the system. One could switch from fast to slow light, with large bandwidth, by tuning the field intensity for the Raman transition between the two hyperfine ground states. 
It is observed that the fast-light associated with the gain for the Raman transition between the two hyperfine ground states could co-propagate with the slow-light associated with loss for the Raman transition between the two hyperfine excited states. Finally, it should be noted that while our scheme is discussed in the context of the sodium $D_1$ transition, the proposed scheme should be applicable to other alkali atoms such as $^7$Li, $^{39,\,41}$K, and $^{87}$Rb) or other molecular systems satisfying the required conditions.

\begin{acknowledgments}
A. H. M. A would like to thank Indian Council for Cultural Relations (ICCR), Government of India and Ministry of Higher Education (MoHE), Egypt for support through a research scholarship.
\end{acknowledgments}

\bibliography{ref}
\end{document}